\documentclass[aps,prl,twocolumn,groupedaddress,showkeys]{revtex4}
\usepackage{graphicx}

\begin{document}
\title{Experiments on the Fermi to Tomonaga-Luttinger liquid transition
in quasi-1D systems }

\author{M. Hilke,$^1$ D.C. Tsui,$^2$ L.N. Pfeiffer$ ^3$ and
K.W. West$ ^3$}

\affiliation{$^1$Dpt. of Physics, McGill University, Montr\'eal,
Canada H3A 2T8,\\$^2$Dpt. of Elect. Eng., Princeton University,
Princeton, New Jersey, 08544,\\$ ^{3}$Bell Laboratories, Lucent
Technologies, Murray Hill,  New Jersey, 07974 }
\thanks{hilke@physics.mcgill.ca}
\date{June 9, 2003}

\begin{abstract}We present experimental results on the tunneling
into the edge of a two dimensional electron gas (2DEG) obtained
with GaAs/AlGaAs cleaved edge overgrown structures. The electronic
properties of the edge of these systems can be described by a
one-dimensional chiral Tomonaga-Luttinger liquid when the filling
factor of the 2DEG is very small. Here we focus on the region
where the Tomonaga-Luttinger liquid breaks down to form a standard
Fermi liquid close to $\nu=1$ and show that we recover a universal
curve, which describes all existing data.
\end{abstract}


\maketitle

In one dimension and in the presence of interactions, a metal can
have a Fermi surface in agreement with Luttinger's theorem
\cite{luttinger}. However, fermionic quasi-particles are no longer
possible and the elementary excitations are replaced by bosonic
charge and spin fluctuations dispersing with different velocities.
Hence, this one-dimensional metal is no longer a Fermi-liquid but
a Tomonaga-Luttinger liquid (TLL) \cite{haldane}. Models
describing one-dimensional interacting Fermions were first
considered by Tomonaga and Luttinger \cite{tomonaga}.

While there are a number of systems, which could  exhibit $TLL$
behavior, Wen \cite{wen} showed that the edge modes of fractional
quantum Hall (FQH) states can be described as chiral $TLL$s. The
chirality is due to the presence of a magnetic field, which forces
the edge states to propagate in one direction. A unique feature of
the {\em chiral} $TLL$ is the absence of back-scattering, i.e., no
localization can occur, which is in stark contrast to the
non-chiral case. For experiments, a key theoretical result is the
existence of power-law correlation functions, which lead to the
vanishing of the momentum distribution function at $k_F$ following
a power-law, i.e., $n(k)\sim|k-k_F|^\alpha$, where $\alpha$ is
related to the interaction strength. As a consequence, the
tunneling current-voltage (I-V) characteristics follows $I\sim
V^\alpha$ and the zero bias conductance follows $\sigma\sim
T^{\alpha-1}$.\cite{wen} For the particular case, where the
filling factor $\nu=1/3$, Wen predicted that $\alpha=3$, hence the
tunneling current and conductivity should vanish like $I\sim V^3$
and $\sigma\sim T^2$ respectively. This is very different from the
Fermi liquid-to-Fermi liquid tunneling which would be ohmic and
independent of temperature.

Following the predictions of Wen \cite{wen} and others
\cite{kane}, several experimental attempts were made in order to
observe this power-law dependence. The first experiments
considered a gate induced constriction to tunnel between two FQH
liquids \cite{proexp,conexp}. Unfortunately, in some cases the
results were consistent with a power-law \cite{proexp} but not in
others \cite{conexp}. This was largely attributed to the
smoothness of the potential barrier causing the possible
reconstruction of the edge and an energy dependent tunneling
barrier. Chang et al. \cite{chang} avoided this problem by growing
a sharp tunneling barrier on the cleaved edge of a two dimensional
electron gas (2DEG). They obtained a good power-law over more than
a decade in voltage to obtain a tunneling exponent ($\alpha\simeq
2.7$ at $\nu=1/3$) close to Wen's prediction.

When moving away from the primary fraction $\nu=1/3$ to
$\nu=p/(2np\pm 1)$ (where p and n are positive integers), the edge
cannot be described anymore by a single LL edge mode but requires
several additional modes, the number and nature of which depends
strongly on the particular fraction and, moreover, the disorder
becomes important because of possible inter-mode scattering. The
overall structure of these states is reviewed in ref. \cite{wen2}.
As a consequence, the recent experimental result from Grayson et
al. \cite{grayson} came as a surprise, because instead of
observing a plateau-like structure between $\nu^{-1}=2$ and 3, as
expected from both the composite fermion theory
\cite{shytov,levitov} and a disordered edge in the hierarchical
model \cite{polchinski}, they observed a linear dependence of the
exponent on the inverse filling factor, $\alpha\simeq\nu^{-1}$.
Recent theories have attempted to account for this behavior using
different approaches \cite{theopro} and are currently under
debate.

\begin{figure}[h]
\begin{center}
\includegraphics[scale=0.3]{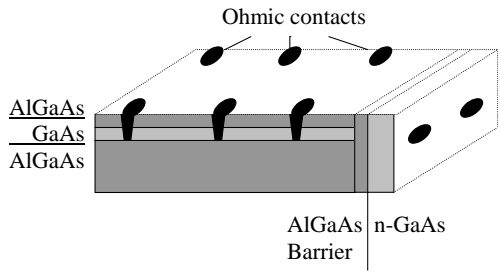}
\caption{ Sketch of the cleaved edge overgrown structure. The
contacts are made by first annealing the Indium 2DEG contacts and
then the shallow 3D contacts. }
\end{center}
\end{figure}

In this article, we are interested in the transition between the
Fermi liquid and the non-Fermi liquid (or TLL) behavior. It is
generally expected that at $\nu=1$ the edge should behave as a
standard Fermi liquid and all theories agree. Away from $\nu=1$
the situation is very different, for both $\nu<1$ and $\nu>1$. We
experimentally probe the edge around this filling factor by using
very high mobility 2DEGs (between $1-30\times 10^6$cm$^{-2}$/Vs),
which all show well pronounced fractional quantum Hall features.
The 2DEG is confined in GaAs/AlGaAs quantum wells. The 2DEG sample
is then placed in the molecular beam epitaxy (MBE) chamber and
cleaved for a subsequent growth along the (110) direction. First
an atomically sharp barrier of (Al$_{x}$Ga$_{1-x}$As) is grown and
then a n-doped GaAs layer \cite{loren}. We fixed the barrier
height to about 200meV, by using $x=0.2$ as barrier material. The
value of the tunnelling resistance can be tuned by changing the
barrier width. We typically used a barrier between 60\AA $ $ and
120\AA $ $ wide. A sketch of the structure is shown in fig.1.

The n-doped GaAs layer plays a crucial role because it defines the
electrostatic potential at the interface between the 2D and 3D.
Indeed, when using a high 3D doping or density the resistance
across the barrier is extremely high ($>G\Omega$), even at room
temperature and independent on the barrier width. We believe that
this is due to important electrostatic depletion. In order to
analyze the 3D density, we alloyed two shallow ohmic contacts
directly on the edge and measured the Shubnikov de Haas
oscillations. Typical curves of these are shown in fig. 2. For
these densities, the zero field tunnelling resistance depends
mainly on the barrier width. A representative list of the
different samples is shown in table I below, with their
corresponding densities.

\begin{figure}[h]
\begin{center}
\includegraphics[scale=0.3]{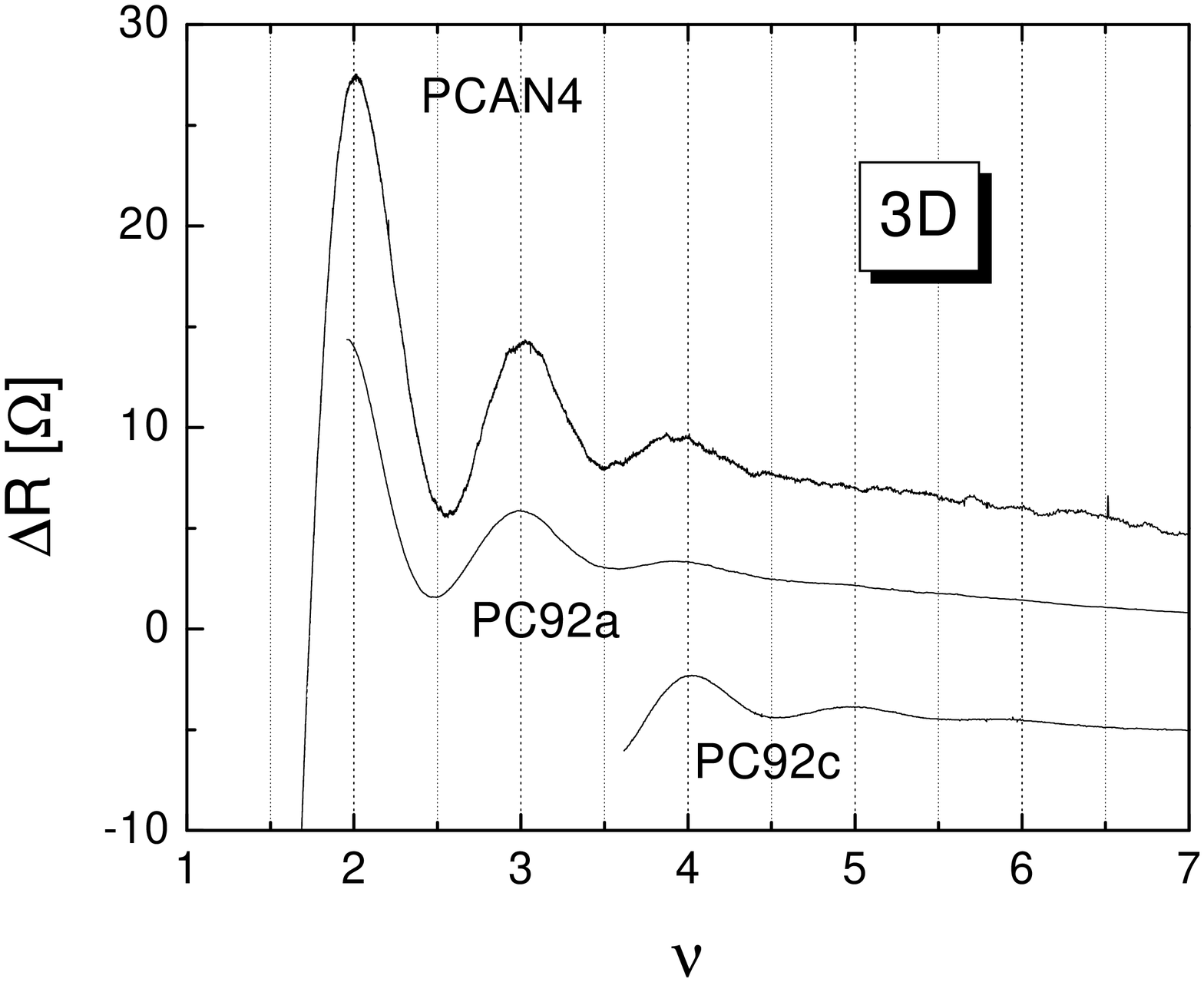}
\caption{  Magnetoresistance measured directly on the edge as a
function of the normalized inverse magnetic field at 40mK. The
extracted densities are shown in the table. }
\end{center}
\end{figure}

\begin{table}
\caption{Sample densities.} \label{t2}
\begin{tabular}{@{\hspace{\tabcolsep}\extracolsep{\fill}}cccc}
\hline Name & 2D Density & 3D Density \\ \hline \textbf{PCS2c2}  &
$1.73\times 10^{11} cm^{-2}$ &
 \\ \textbf{PC92a}  & $3.35\times 10^{11} cm^{-2}$ &
$3.25\times 10^{17} cm^{-3}$ \\ \textbf{PC92c}  & $3.35\times
10^{11} cm^{-2}$ & $8.5\times 10^{17} cm^{-3}$\\ \textbf{PCAN2} &
$2.2\times 10^{11} cm^{-2}$ & $4.5\times 10^{17} cm^{-3}$\\
\textbf{PCAN4}  & $2\times 10^{11} cm^{-2}$ & $4.5\times 10^{17}
cm^{-3}$\\ \textbf{PCL3}  & $1.15\times 10^{11} cm^{-2}$ &
\\ \hline
\end{tabular}
\end{table}

In fig. 3 we show a typical four terminal magnetoresistance trace
across the barrier. In this particular sample the tunnelling
resistance is very low at B=0, which allows us to identify all the
quantum Hall features typical of a high mobility 2DEG. Above 6T
the tunnelling barrier becomes very resistive and we observe an
exponential increase in tunnelling resistance with B. This is
generic to all samples. The main difference between samples is the
tunnelling resistance at $B=0$. The overall magnetic field
dependence is dominated by an exponential increase of the
resistance as a function of $B$, which is due both to the
suppression of momentum conservation and to the finite extent of
the wave-function given by the magnetic length.

\begin{figure}[h]
\begin{center}
\includegraphics[scale=0.3]{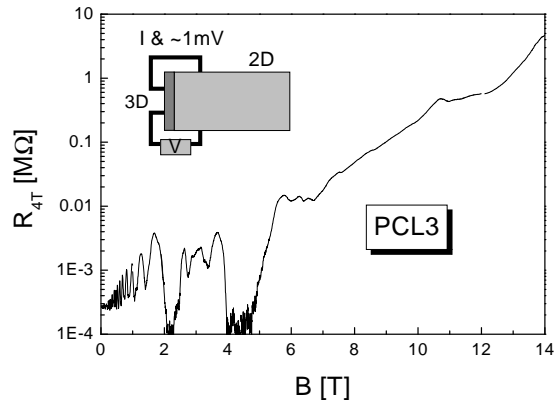}
\caption{ The magnetoresistance of sample PCL3, measured using a
four-terminal configuration, as a function of magnetic field. The
data was obtained by applying 1mV between the 3D and 2D and
measuring the current with a source-meter. The four-terminal
voltage drop is measured by using a voltage amplifier with a high
input impedance. }
\end{center}
\end{figure}

We measured the current-voltage (I-V) characteristics at different
magnetic fields. Overall, the I-V's are essentially linear at low
fields and become increasingly non-linear at higher fields. This
non-linearity does not depend on the tunnelling resistance since
the behavior is very similar in samples who have very different
tunnelling resistances. We found that in all samples the filling
factor is the relevant parameter, which determines whether we have
a linear or a non-linear I-V. Following the analysis of
refs.\cite{chang,grayson,chang2,hilke} we extract $\alpha$ from
the power-law of the I-V's, using two different methods. We can
either use the temperature dependence of the I-V traces or
directly the non-linearity in the I-V. A typical temperature
dependent I-V curve is shown in fig. 4, in which the filling
factor is $\nu=2.7$.

\begin{figure}[h]
\begin{center}
\includegraphics[scale=0.3]{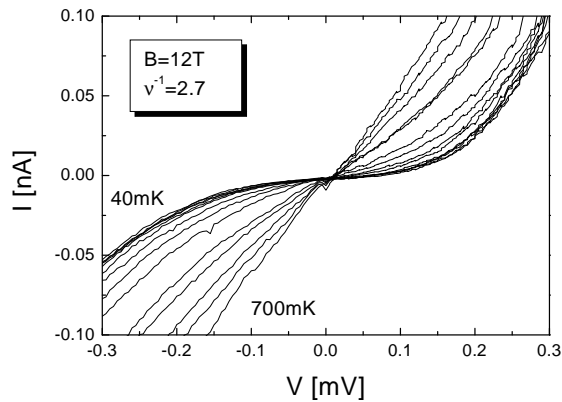}
\caption{Temperature dependence of the I-V curves at 12T,
corresponding to a filling factor of $\nu^{-1}=2.7$. This set of
curves is taken with sample PCL3. }
\end{center}
\end{figure}

From the slopes of the $I-V$ traces we can extract the zero bias
conductance as a function of temperature, which is plotted in fig.
5. Above 100mK the conductance clearly follows a power-law
dependence over one decade in temperature. When fitting the data
with a power law in that range we obtain a power of 1.64$\pm0.03$.
The saturation below 100mK is a consequence of the finite input
resistance of our voltmeter and not related to the sample. We can
avoid this saturation by using a two-terminal configuration for
more resistive samples, since in that case the two-terminal
contribution from the 2DEG  becomes negligible.

\begin{figure}[h]
\begin{center}
\includegraphics[scale=0.3]{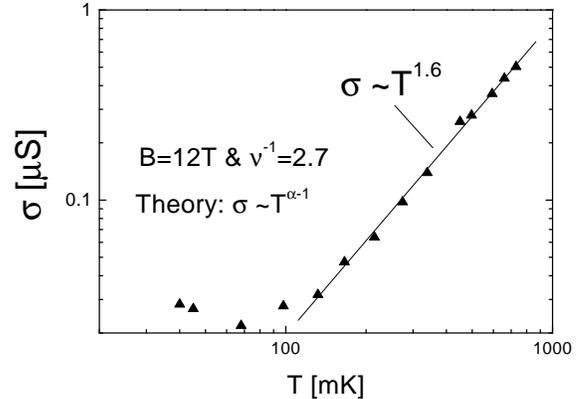}
\caption{Temperature dependence of the zero bias conductance at
12T obtained from the slopes in fig. 4. }
\end{center}
\end{figure}

In fig. 6 the I-V characteristic is drawn on a log-log scale at
base temperature (40mK). The data follows a power-law over more
than a decade in voltage before it starts to deviate significantly
from a power-law. We obtain a power of 2.45$\pm 0.01$, when
fitting the data below 1mV. Following Wen's \cite{wen} argument,
this would lead to $\alpha=2.45$. This value is very close to the
one obtained from the temperature dependence, i.e., $\alpha=
1+1.64$.

\begin{figure}[h]
\begin{center}
\includegraphics[scale=0.3]{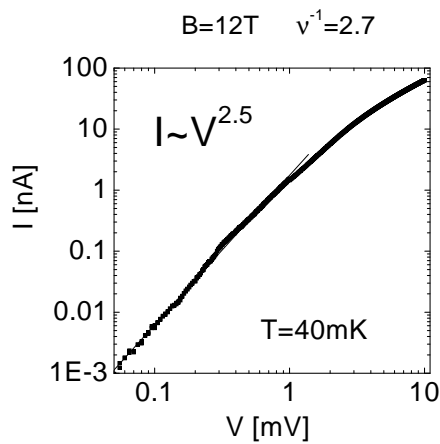}
\caption{The I-V characteristics of sample PCL3 at 12T and 40mK on
a $log-log$ scale. }
\end{center}
\end{figure}

We have repeated this procedure for different magnetic fields and
different samples. In fig. 7 we have compiled all the values of
$\alpha$ as a function of the inverse 2DEG filling factor.

\begin{figure}[h]
\begin{center}
\includegraphics[scale=0.3]{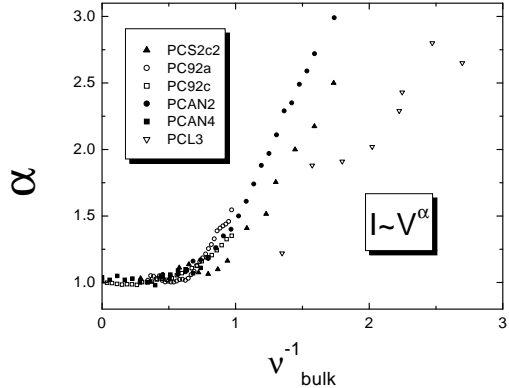}
\caption{Exponent $\alpha$ extracted from the power-law part of
the I-V curves as a function of $B$ and inverse filling factor
$\nu^{-1}$ for samples in table I. The errors are within the size
of the dots.}
\end{center}
\end{figure}

It is striking that all samples behave qualitatively in a very
similar way. Indeed, $\alpha\simeq 1$ for low magnetic fields (low
inverse filling factor) and all the way up to a field
corresponding to a filling factor between 1 and 1.5. When
Increasing the field further there is a transition to a non-linear
regime, in which the exponent increases linearly with the inverse
filling factor. The field at which this transition occurs
corresponds to the {\bf Fermi liquid to TLL
transition}.\cite{hilke} Unexpectedly, this transition occurs at a
filling factor larger than one, which implies that at $\nu=1$ the
system is {\bf not} a Fermi liquid. In addition, this behavior
questions the universal nature of the TLL at the edge of a FQHE
system, since $\alpha$ depends on sample parameters.

But; when rescaling the filling factor of all our samples with a
value such that $\nu=\alpha=1$ at the linear intercept, we can
collapse all our data onto a single curve shown in fig. 8. This
rescaling factor depends on the sample growth structure but is
identical when using two samples from the same growth. We can
perform the same operation on Chang and Grayson's earlier data,
\cite{chang,grayson} by multiplying the filling factor by 1.4. (We
haven't included the recent data from ref. \cite{chang2} since the
data cannot be extrapolated consistently all the way down to
$\alpha=1$ because most of their data is clustered around
$\alpha=3$.) Here again all the data points fall on the same
curve. This is very remarkable, considering that the different
data points stem from very different sample growth parameters.
Indeed, the barrier width varies between 60\AA $ $ and 225\AA $ $
and the $Al$ content of the barrier between 10\% and 20\%. The
different 2D and 3D densities cover a factor 3 in range.

\begin{figure}[h]
\begin{center}
\vspace*{2cm}
\includegraphics[scale=0.3]{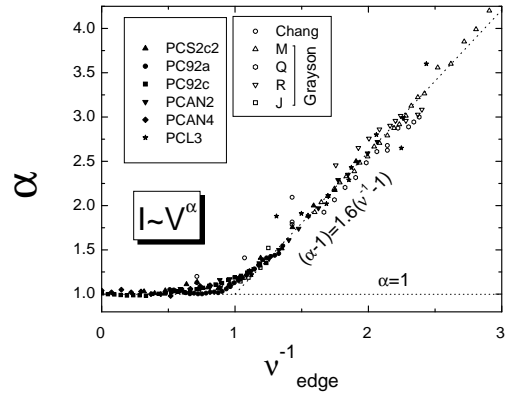}
\vspace*{-2cm} \caption{The chiral TLL exponent $\alpha$ as a
function of the scaled inverse filling factor. The scaling factor
is chosen in such a way that the linear intercepts (dotted lines)
meet at $\nu=\alpha=1$. Our data is presented in filled symbols
and earlier data by Chang, Grayson and co-workers
\cite{chang,grayson} is presented in open symbols.}
\end{center}
\end{figure}

After rescaling the filling factor, we can describe all the data
with the generic form $(\alpha-1)=1.6(\nu_{edge}^{-1}-1)\pm 0.2$,
when $\nu_{edge}<1$. We named this rescaled filling factor
$\nu_{edge}$, since the edge filling factor of our 2DEG might be
different from the bulk filling factor. The exact dependence of
this rescaling factor on sample properties, such as the 2D and 3D
densities is currently under investigation. It is interesting to
note that the extrapolation of our curve to $\alpha=3$ is
$\nu^{-1}\simeq 2+1/4$, which is slightly off from most existing
theories.\cite{wen,kane,wen2,shytov,levitov,polchinski,theopro,tsiper}
Moreover, recent theories \cite{levitov,tsiper} suggest that the
2DEG density is not monotonic and even oscillates as one
approaches the edge. This could be indicative of a complex
dependence of the effective edge filling factor $\nu_{edge}$ on
various sample parameters.

In summary, we have presented experimental results on the
Tomonaga-Luttinger to Fermi liquid transition at the edge of a
2DEG system close to $\nu=1$. Although the filling factor of the
transition is different for different sample structures, we are
able to collapse all the data onto a single curve if we rescale
the filling factor using a single parameter, hence recovering the
notion of a universal behavior in the chiral TLL.

We acknowledge support from NSF, NSERC and FCAR

\end{document}